\documentclass{article}
\usepackage{amsmath}
\usepackage{listings}
\usepackage{microtype}
\usepackage{libertine} % Use Libertine font
\usepackage[T1]{fontenc} % Use T1 font encoding
\usepackage[utf8]{inputenc} % Use UTF-8 encoding
\usepackage{array}
\usepackage{tabularx}
\usepackage [english]{babel}
\usepackage [autostyle, english = american]{csquotes}
\usepackage{authblk}  % Package for author affiliations
\usepackage{abstract}
\usepackage{hyperref}
\usepackage{breakurl}
\usepackage[numbers]{natbib}

\newcommand{\keywordsname}{Keywords}  % Define the name for the keywords section
\newcommand{\keywords}[1]{% Define the command for the keywords
  \begin{flushleft}
    \small\textbf{\keywordsname:} #1
  \end{flushleft}
}

\MakeOuterQuote{"}

\AtBeginDocument{%
  }

\begin{document}

\title{Unsafe Impedance: Safe Languages and Safe by Design Software}

%%\author{Ben Trovato}
%%\authornote{Both authors contributed equally to this research.}
%%\email{trovato@corporation.com}
%%\orcid{1234-5678-9012}
%%\author{G.K.M. Tobin}
%%\authornotemark[1]
%%\email{webmaster@marysville-ohio.com}
%%\affiliation{%
%%  \institution{Institute for Clarity in Documentation}
%%  \city{Dublin}
%%  \state{Ohio}
%%  \country{USA}
%%}

\author[1]{Lee Barney}
\author[2]{Adolfo Neto}

\affil[1]{Brigham Young University-Idaho, Rexburg, USA \\
          \texttt{barneyl@byui.edu}}
\affil[2]{Universidade Tecnológica Federal do Paraná, Curitiba, Brazil \\
          \texttt{adolfo@utfpr.edu.br}}

\newcommand\change[1]{{#1}}


\maketitle

\begin{abstract}
In December 2023, security agencies from five countries in North America, Europe, and the south Pacific produced a document encouraging senior executives in all software producing organizations to take responsibility for and oversight of the security of the software their organizations produce. 
\change{In February 2024, the White House released a cybersecurity outline, highlighting the December document.} In this work we review the safe languages listed in these documents, and compare the safety of those languages with Erlang and Elixir, two BEAM languages. 

These security agencies' declaration of some languages as safe is necessary but insufficient to make wise decisions regarding what language to use when creating code. We propose an additional way of looking at languages and the ease with which unsafe code can be written and used. We call this new perspective \textit{unsafe impedance}. We then go on to use unsafe impedance to examine nine languages that are considered to be safe. 
\change{Finally, we suggest that business processes include what we refer to as an Unsafe Acceptance Process.}
This Unsafe Acceptance Process can be used as part of the memory safe roadmaps suggested by these agencies.  Unsafe Acceptance Processes can aid organizations in their production of safe by design software.
\end{abstract}

\keywords{Memory Safe Languages, Functional Programming, Secure By Design, Secure By Default}

\section{Introduction}
Computers connected to networks exist in a dangerous space. From viruses to bots \cite{fedele2022dangerous}, bad actors are constantly looking for weaknesses they can exploit. In response to this, a group of national security agencies stated that C-Suite executives have a responsibility along with technical experts to reduce the attack surface of software over which they have responsibility \cite{cybersecurity2023case}. The proposal they make is to migrate systems from being written in what they refer to as memory unsafe languages to Memory Safe Languages (MSLs).

In their mitigations, these agencies state, 

"\textit{Even the most experienced developers write bugs that can
introduce significant vulnerabilities. Training should be a bridge while an organization
implements more robust technical controls, such as memory safe languages.}" \cite{cybersecurity2023case}

They also suggest several other activities, from implementing and enforcing coding guidelines to hardware changes, that may mitigate memory based Common Vulnerability and Exposure (CVE) Types when using memory unsafe languages. All of this is done as a preamble to the major suggestions of the publication, the use of and transition to MSLs. 

As part of a move to a more secure future, these agencies urge all software manufacturers, not just those that sell or give away software, to produce and publish "memory safe roadmaps" indicating how they are going to take ownership of the "security outcomes" \cite{cybersecurity2023case} of their software and develop secure products. All of this to promulgate understanding amongst all producers of software that "the software industry needs more secure products, not more security products." \cite{secureByDesignShifting}

All of the MSLs mentioned in the articles published by these security agencies allow memory unsafe code to be directly written in the language or loaded from libraries. If unsafe code is written or loaded to produce a product, the product is then is written using unsafe code. This implies that moving to MSLs is necessary but insufficient. Unsafe impedance, as we shall define it, is an additional way to aid technical experts and C-suite executives to choose languages and build their memory safe roadmaps. It is also a way to aid creators of MSLs as they contemplate the design of their new language.

The contributions of this paper are:
\begin{itemize}
\item Introduction of the concept of "unsafe impedance" as a novel perspective for evaluating the safety of programming languages in the context of software security.
\item Review and comparison of safe languages listed in cybersecurity documents from security agencies with Erlang and Elixir, two BEAM languages.
\item Proposal for an Unsafe Acceptance Process (UAP) to enhance software security by evaluating the necessity and risks associated with incorporating Native Implemented Functions (NIFs) in Erlang or Elixir applications.
\end{itemize}
\section{Secure by Design and Secure by Default}
In 2023, thirteen security agencies from various nations around the world described technology products that are "secure by design" and "secure by default" \cite{secureByDesignShifting}. They define secure by design as it relates to technology products as meaning these products "\textit{are built in a way
that reasonably protects against malicious cyber actors successfully
gaining access to devices, data, and connected infrastructure}"  \cite{secureByDesignShifting}.  

These same agencies define secure by default as "\textit{products are resilient against prevalent
exploitation techniques out of the box without added charge}"  \cite{secureByDesignShifting}.

Considering a programming language as a product, we build upon these definitions and define secure by default languages.

\textit{Languages are secure by default if they are created in a way that protects against the production of all types of Common Vulnerability and Exposures (CVEs).}

We will focus our assessments of languages in regard to being secure by default with regard to the languages' ability to create memory CVEs. If a language is not secure by default, we define it to be \textit{unsecure by default}.

While all categorization schemes are flawed, we categorize memory CVEs into spacial and temporal groupings. Spacial memory CVEs being those where memory locations and their use cause vulnerabilities and exposures and temporal memory CVEs being those where time differences cause memory vulnerabilities and exposures.

\begin{table}[h!]
    \caption{Examples of Spacial and Temporal Memory CVEs}
    \centering
    \begin{tabularx}{\columnwidth}{|>{\raggedright\arraybackslash}X|>{\raggedright\arraybackslash}X|}
        \hline
        \textbf{Spacial Memory CVEs} & \textbf{Temporal Memory CVEs} \\
        \hline
        Buffer overflow \cite{piromsopa2006buffer}, & Use after free \cite{Yason2013}, \\
        Buffer underflow \cite{gil2018there}, & Double free \cite{caballero2012undangle}, \\
        Array index out-of-bounds \cite{cybersecurity2023case}, & Dangling pointers \cite{caballero2012undangle}, \\
        Pointer arithmetic errors \cite{bojanova2022data}, & Memory leaks \cite{sun2018projection}, \\
        Uninitialized memory reads \cite{bruening2011practical} & Data race conditions \cite{alsharhan2024parallel} \\
        \hline
    \end{tabularx}
    \label{tab:memory_cves}
\end{table}

The CVEs in Table \ref{tab:memory_cves} and their mitigations are commonly taught in undergraduate computer science courses, and commonly dealt with when hardening software against attack.
\section{Memory Safe Languages and Unsafe Code}
In their December 2023 report \cite{cybersecurity2023case}, eight security agencies list six languages as being \textit{memory safe}, C\#, Go, Java, Python, Rust, and Swift. However, none of these languages are \textit{secure by default}. Each of these languages, regardless of their safety declarations, allow unsafe code to be written and used within the languages' safe code. It is true, however, that each language has its own requirements that are enforced when unsafe code is used. 

We define the difficulty experienced by programmers when when complying with these requirements  as \textit{unsafe impedance}. Choosing languages with a high unsafe impedance as part of the product design makes it easier to claim that the product is secure by design.

For each language listed in the December 2023 report, we give one or more examples of how to use unsafe code. The examples are not intended to indicate how to safely use unsafe code in that language. Neither do we claim these code snippets to be common uses of unsafe code in these languages since the common uses of unsafe code in these languages \change{vary} widely. The purpose of these code snippets is to allow the reader to assess the amount of unsafe impedance programmers in each language experience. 

Also, we are not implying that any programmer or engineer would purposefully write the code in these snippets.

We define languages as having \textit{no unsafe impedance} when they have no syntactical or other impediments to writing unsafe code. Commonly known examples of this language grouping are C and C++. Also, these two languages are unsecure by default.

Languages that have few  syntactical or other impediments to writing unsafe code we define as having \textit{low \allowbreak unsafe \allowbreak impedance}.

Languages that have many syntactical or other impediments to writing unsafe code we define as having \textit{high unsafe impedance}. 

Languages that do not allow any writing, loading, or use of unsafe code we define as having \textit{infinite unsafe impedance}. At this time we offer no rubric to rank languages with regards to unsafe impedance.

\subsection{C\#}

The C\# \cite{hejlsberg2010} language uses a static method of the Marshall class to allocate an unsafe array of specified types, \textbf{int} in the snippet below. Notice that \textbf{Marshal.AllocHGlobal} does not initialize the allocated memory. Any old data stored in the memory remains. This can cause an uninitialized memory read spacial memory issue unless the applications' programmers are sufficiently experienced so that they are aware of the need to write extra code to initialize the memory with some set of default values.

Additionally, there is no verbiage or any other indicator stating that this code, and code using the results of this code, are unsafe. Instead, the programmers creating this code and those who latter read and debug this code are required to gather the knowledge that this code is unsafe from external sources.

\begin{lstlisting}[language=C]
IntPtr pointer = Marshal.AllocHGlobal(5 * sizeof(int));
\end{lstlisting}

When memory is allocated using \textbf{Marshal.AllocHGlobal}, it is required of the programmers to use the unsafe static method, \textbf{Marshal.FreeHGlobal} \change{(see code snippet below)}. If this function is called at the wrong time, the application will experience temporal memory issues such as a dangling pointers, use after free, and double free. if the programmers fail to call this function in an appropriate location in their codebase, the memory leak temporal memory issue is created. In non-simple applications it can be difficult for programmers to know the correct or even a good location to call \textbf{Marshal.FreeHGlobal}.
\begin{lstlisting}[language=C]
Marshal.FreeHGlobal(pointer);
\end{lstlisting}

In addition to the allocate and free issues already described, unsafe pointers can be created, manipulated, and possibly misused. For these reasons, \change{\textit{we classify C\# as being unsecure by default}}.

Also, C\# allows programmers to execute unsafe code from within their safe code with no syntactical or other indication that the code they are writing is unsafe. For this reason, \change{\textit{we would categorize C\# as having no unsafe impedance}}. 

\subsection{Go}
In Go \cite{donovan2015}, an unsafe pointer can be created with no visual indicator that the pointer is unsafe.
\begin{lstlisting}[language=Go]
var ptrToUint32 *uint32 = nil
fmt.Println("Value:", *ptrToUint32)
\end{lstlisting}
As in C, any nil pointer in Go can be dereferenced causing undefined behavior. Since Go allows the creation, manipulation, and possible misuse of pointers, \change{\textit{we classify Go as unsecure by default}}.

It is true that users of the Go language can choose to include the name of the \textbf{unsafe} package from Go's standard library in their code. This is a visual indicator that the result is unsafe to use.
\begin{lstlisting}[language=Go]
var num uint32 = 200
var ptrToUint32 *uint32 = (*uint32)(unsafe.Pointer(&num))
*ptrToUint32 = 300
\end{lstlisting}

However, the use of this \textbf{unsafe} indicator is not required.

\begin{lstlisting}[language=Go]
var num uint32 = 200
var ptrToUint32 *uint32 = &num
*ptrToUint32 = 300
\end{lstlisting}

\change{There is no requirement enforced by the Go language's syntax to indicate that a variable is unsafe.}
For programmers with limited understanding of unsafe code's behaviors, this lack of visual indicators can lead to security vulnerabilities. For this reason \change{\textit{we would categorize Go as having low unsafe impedance}}.

\subsection{Java}
In Java \cite{ullenboom2022}, any library that can be accessed using a C-style header can be loaded and run via Java Native Interface (JNI) functions. There are no constraints on what unsafe code can do in those non-Java native functions.  Anything that is unsafe in that library becomes a hidden unsafe behavior in the encapsulating Java code which can then be used in any number of places in the Java application.

Here is the Java side of the JNI relationship of a simple function that adds two ints.
\begin{lstlisting}[language=Java]
public class AdditionExample {

    // Declare the native method
    public native int add(int x, int y);

    // Load the native library
    static {
        System.loadLibrary("addition");
    }
}
\end{lstlisting}
Notice there is no explicit visual indicator that unsafe code may be executed. The \textbf{native} keyword is insufficient in that a novice reading the code is not directly told the code may be unsafe.

While it is true that several steps are required to complete the library, the impedance generated by these steps is reduced by the Java system providing tools to make it easier to execute unsafe JNI functions. For this reason, 
\change{\textit{we also are categorizing JNI interactions as having a low unsafe impedance.}}

Foreign Function and Memory (FFM) has been introduced as part of Project Panama \cite{openJDKPanama} and is scheduled for release as part of Java 22. FFM allows direct direct memory manipulation and is designed to replace JNI. Based on the FFM documentation, FFM uses sandboxes, referred to as arenas, to try to contain and constrain the behaviors of unsafe code executed via the FFM API. In a presentation regarding FFM, it was stated that FFM attempts to "find a balance" \cite{Cimadamore2023Panama} between safety and flexibility. 

The lack of visual unsafe cues in code examples using FFM and the balance between safety and ease of use \change{\textit{lead us to categorize this type of Java code as having a low unsafe impedance.}} 
\change{\textit{We also categorize Java as being unsecure by default because of the safety/flexibility tradeoff of FFM.}}

\subsection{Python}
Python \cite{summerfield2009} programmers can use the \textbf{ctypes} library to interact with dynamically loaded libraries. These libraries can be written in any language that can produce libraries "which export functions using the standard cdecl calling convention" \cite{pythonCtypes}. This includes languages such as C, Rust, and Swift.

Compilation and testing of these libraries is done external to the Python language and its standard toolset. There are non-standard toolsets designed to ease the creation and integration of these compiled libraries.

\change{
When a Python application loads a dynamically linked library using \textbf{ctypes.CDLL}, the library uses the memory space of the Python REPL.
}
This gives any security vulnerability access to any data and code in the REPL's memory space. 
%This means there would be no additional safety for code and data run in the REPL when compared to if the code was written in the language of the library. 
\change{This means there is no additional safety for code and data run in the REPL compared to if the code were written in the language of the library.}
%It is true that if the library was loaded by an additional Python REPL that this new REPL memory space would only include the data used and created by the loaded library, but this does not produce any added safety.
\change{
    It is true that if the library were loaded by an additional Python REPL, the new REPL's memory space would only include the data used and created by the loaded library. However, this does not provide any additional safety.}

Notice that the use of the dynamically loaded library in the example below does not include any direct indicators that the code being used is unsafe.

\begin{lstlisting}[language=Python]
import ctypes

libc = ctypes.CDLL('libc.dylib')

libc.malloc.argtypes = [ctypes.c_size_t]
libc.malloc.restype = ctypes.c_void_p
libc.free.argtypes = [ctypes.c_void_p]

num_elements = 10
element_size = ctypes.sizeof(ctypes.c_int)
array_ptr = libc.malloc(num_elements * element_size)
\end{lstlisting}
% deallocating an array used with external libraries
% \begin{lstlisting}[language=Python]
% libc.free(array_ptr)
% \end{lstlisting}

Python's wrapping of unsafe arrays in Python types for access and modification does add some security when manipulating them within Python. However, this does not preclude their misuse within the library itself. Untrustworthy actors could still leverage weaknesses in the library.

Because Python requires unsafe code to be compiled outside of its language, runtime, and tools, 
\change{\textit{we claim Python has a high unsafe impedance.}} 
This, however, is tempered by the lack of indicators for unaware and unknowledgeable programmers or engineers that the code being executed is unsafe.

A version of Python with an infinite unsafe impedance could be created. It would require that no externally compiled code would be allowed to be loaded. This version of Python would, unfortunately, be unavailable for use by the machine learning community since a great deal of the code used by Python machine learning modules is found in dynamically loaded libraries written in unsafe languages.

\change{\textit{We do categorize Python as a secure by default language as long as no unsafe code is loaded and used.}}

\subsection{Rust}
Rust \cite{klabnik2010} includes the keyword \textbf{unsafe}. It is used along with scope indicators to specify a block of code where all Rust's safety rules are ignored by the compiler. Use of the unsafe keyword and code blocks does not add any additional safety to the code or the applications. It is strictly there to indicate to programmers and engineers that CVE's can be executed in the blocks.

Therefore, any unsafe behavior such as use-after-free, dangling pointers, and the other CVEs are purposefully placed inside of these blocks. The intention is that by localizing places where CVEs can exist, they will be easier to manage. These unsafe code blocks also are used to leverage code written in unsafe languages. 

It is common to hide the use of unsafe code in Rust by writing functions that show themselves as being safe and following the compiler's safety rules yet those functions contain unsafe code blocks. This is so common that when discussing using libraries written in unsafe languages, the Rust documentation states that one of these unsafe libraries "can choose to expose only the safe, high-level interface and hide the unsafe internal details" \cite{rustFFI}.

Notice the use of unsafe Rust code in this example. Unsafe behaviors are not limited to libraries written in other languages. Rust itself eases the creation of unsafe code through its own syntax.
\begin{lstlisting}[language=Python]%Rust is not a recognized lstlisting language

use std::alloc::{alloc, dealloc, Layout};
 unsafe {
        let layout = Layout::new::<u16>();
        let ptr = alloc(layout);
}   
.
.
.
 unsafe {
        dealloc(ptr, layout);
    }
\end{lstlisting}

When a Rust application loads a library, the library shares the same memory space with the rest of the application. This causes any Rust application that loads libraries to have the same kind of vulnerabilities as a Python application with loaded unsafe libraries. Because of this and the inclusion of Rust syntax for creating unsafe code, \change{\textit{we categorize Rust as having low unsafe impedance.}} The unsafe keyword and the requirement that the \textbf{mut} keyword be used for mutable variables  would, in our opinion, place Rust in the upper portion of low unsafe impedance category.

Because of Rust's ability to create, manipulate, and possibly misuse pointers, \change{\textit{we categorize Rust as being an unsecure by default language.}}

\subsection{Swift}
When a Swift \cite{swiftBook} programmer is considering writing unsafe Swift code, Swift's syntax enforces the use of an unsafe indicator. This is found in the names of the initialization functions for \textbf{UnsafePointer}, \textbf{UnsafeMutablePointer}, and \textbf{UnsafeRawPointer}. These types map to various C-style pointers and, when included in Swift code, are often used to increase speed or interact with libraries in other languages. The Swift documentation states that  "Swift imports any function declared in a C header as a Swift global function" \cite{swiftImportedCFunctions}. Swift also enables interactions with C macros \cite{swiftImportedCFunctions}, structures, and unions \cite{swiftImportedCFunctions}.

The C header declaration implies that any library written in any language that can expose its functionality as if it was written in C is interoperable with Swift. Like Rust and Python, any such library shares the same memory space as the rest of the Swift application. As with those languages, this opens up the possibility of bad actors gaining access to data in the application and executing nefarious code within the Swift portion of the application.

Below is an example of unsafe code written in Swift.
\begin{lstlisting}[language=Swift]
var pointer = UnsafeMutablePointer<UInt16> .allocate(capacity: count)
.
.
.
pointer.deinitialize(count: count)
pointer.deallocate()
\end{lstlisting}

The first line indicates that the pointer created is unsafe. This unsafe code can be wrapped in Swift code that appears to be safe, hiding unsafe code in what appears to be safe code. Swift's use of unsafe in the initialization functions for unsafe types 
\change{\textit{in our opinion places Swift in the upper portion of low unsafe impedance category.}}

Swift's native ability to create, manipulate, and possibly misuse pointers 
\change{\textit{causes us to categorize Swift as an unsecure by default language.}}

\subsection{Erlang, Elixir, and other BEAM Languages}
Erlang \cite{armstrong2007}, Elixir \cite{thomas2018}, and other programming languages run on the BEAM virtual machine \cite{virdingBeam} which is often referred to as 'the BEAM'. The BEAM allows unsafe functions to be loaded into its memory and execution space. Like Rust, Swift, and the other languages described as safe in the December 2023 report \cite{cybersecurity2023case}, this means any unsafe, loaded code may be leveraged by bad actors and give the bad actors access to data in the application and executing nefarious code within the BEAM.

Because of this, the Erlang documentation includes this warning, "An erroneously implemented native function can cause a VM internal state inconsistency, which can cause a crash of the VM, or miscellaneous misbehaviors of the VM at any point after the call to the native function" \cite{erlNIF}. 

Unlike Swift, the libraries containing unsafe code, one or more Native Interface Functions (NIFs), have to include a specific header and deal with some Erlang terminology. Below is a small example.
 
\begin{lstlisting}[language=c]
// example_nif.c
#include "erl_nif.h"

static ERL_NIF_TERM create_array(ErlNifEnv* env, int argc, const ERL_NIF_TERM argv[]) {
    // Example function that allocates an array and returns a pointer as an integer
    unsigned int* array = malloc(sizeof(unsigned int) * 10);
    if(array == NULL) {
        return enif_make_badarg(env);
    }
    // Just an example: initializing array with arbitrary values
    for(int i = 0; i < 10; i++) {
        array[i] = i;
    }
    return enif_make_uint64(env, (ErlNifUInt64)array);
}

static ERL_NIF_TERM free_array(ErlNifEnv* env, int argc, const ERL_NIF_TERM argv[]) {
    // Expects a pointer as an unsigned integer
    ErlNifUInt64 ptr;
    if(!enif_get_uint64(env, argv[0], &ptr)) {
        return enif_make_badarg(env);
    }
    free((void*)ptr);
    return enif_make_atom(env, "ok");
}

static ErlNifFunc nif_funcs[] = {
    {"create_array", 0, create_array},
    {"free_array", 1, free_array}
};

ERL_NIF_INIT(example_nif, nif_funcs, NULL, NULL, NULL, NULL)

\end{lstlisting}

It is possible to use the \textbf{create\_array} function from Erlang code without knowing it is unsafe. BEAM languages such as Erlang and Elixir can hide unsafe code in much the same way Rust, Swift, and the other languages mentioned do. 

The Erlang and Elixir languages are functional, declarative, and "secure by default"  \cite{secureByDesignShifting}. As part of this secure by default approach, the creators of Erlang and Elixir have not included the ability to create or use pointers in the languages, unlike Rust and Swift. Therefore nine of the ten common CVEs listed in Table \ref{tab:memory_cves} are irrelevant assuming no NIFs are used in the creation of the Erlang library, application, or system. The tenth CVE from Table \ref{tab:memory_cves}, data race conditions, cannot occur when using variables since all Erlang and Elixir variables, tuples, lists, maps, etc. are immutable.

This is not to say that there are no poor programming practices in Erlang and Elixir that can be problematic. The Erlang Ecosystem Foundation's Security Working Group provides guidance for avoiding these poor practices \cite{secureCodingErlef}. As an example, in large, long-running systems, it is possible to exceed the the number of atoms available since atoms are not garbage collected. The number of available atoms for a node is determined at startup. The default value is 1,048,576, which can by increased by using the \textbf{+t} flag. 

It is unwise to accept large amounts of arbitrary data that is converted to atoms.  When data must be converted to atoms, the application of other interventions such as using \textbf{list\_to\_existing\_atom/1} and then using \textbf{list\_to\_atom/1} if \textbf{list\_to\_existing\_atom/1} fails is indicated. This type of practice preserves the limited atom table resource.

The awkwardness of the creation of NIFs, the requirement that they be compiled outside of the standard build environment, and the inherent safety of Erlang and Elixir code \change{\textit{in our opinion places Erlang and Elixir in the high unsafe impedance category. We believe it is in the lower end of this category due to the ability to hide the execution of unsafe code.}}

We do categorize Erlang and Elixir as a secure by default languages as long as no unsafe code is loaded and used.

\begin{table}[htbp]
\caption{The Unsafe Impedance and Security Defaults for Several Languages}
\centering
\begin{tabular}{p{1.1cm}|>{\raggedright\arraybackslash}p{1.2cm}|>{\raggedright\arraybackslash}p{1.2cm}|>{\raggedright\arraybackslash}p{1.2cm}|>{\raggedright\arraybackslash}p{1.2cm}|}
\cline{2-5} % Horizontal line from column 2 to 5
\multicolumn{1}{c|}{} & \multicolumn{4}{|c|}{\textbf{Unsafe Impedance}} \\
\cline{2-5}
\textbf{} & \textbf{none} & \textbf{low} & \textbf{high} & \textbf{infinite} \\
\hline
\textbf{Secure by Default} &  &  & Python, Erlang, Elixir &  \\
\hline
\textbf{Insecure by Default} & C, C++,C\# & Go, Java, Swift, Rust & & \\
\hline
\end{tabular}
\label{tab:secureImpedance table}
\end{table}

\section{Secure By Design Software Using Erlang and Elixir}\label{sec:secureErlangElixir}

To achieve secure by default behavior in an application written using secure by default languages such as Erlang and Elixir, business processes must be implemented and enforced that strongly discourage the use of NIFs. We propose that any group producing software in any safe by default language create what we refer to as an \textit{Unsafe Acceptance Process} (UAP). The purpose of any UAP is to increase unsafe impedance. The UAP must impose a significant barrier to the loading and use of potentially unsafe code. 

We also propose that to be of most use, any UAP must include a required, measurable proof that only a NIF can solve the problem presented.

We propose this proof should include at least these items in a proposal to include a NIF in an application:
\begin{enumerate}
\item a statement that by the lack of a NIF implementation one or more existing consumers/users of the software are being damaged and how,
\item a statement of the security risks any NIF brings, along with a statement of the potential security risks posed by the NIFs code,
\item the source code for the proposed NIF,
\item the source code for unit tests, when unit testing is possible, or other extensive proofs that the NIF exhibits no unsafe behavior for all conceivable edge cases,
\item measured speed increases presented by using the NIF, and
\item an alternative solution, if possible, in Erlang or Elixir that improves on the current solution, but is insufficient.
\end{enumerate} 

As part of any UAP developed by software producing organizations, no programmer or software engineer should be able to independently add potentially unsafe code to a library, application, or system. The NIF proposal should be evaluated by technological and business persons responsible for the software. Such an evaluation must be skeptical, and adverse to the addition of any NIF in its initial perspective. 
%This implies addition of a NIF must be supported by compelling and overwhelming evidence of the need for the NIF.
\change{
This implies that the addition of a NIF must be supported by compelling and overwhelming evidence of its necessity.
}

In our opinion, if a business process like the one described here is implemented for Erlang or Elixir applications, the resulting code would fall into the middle of the high unsafe impedance classification.

\section{Conclusions}
It is possible for any system written in safe languages to use unsafe code and unwittingly expose the system to attack. For some of the languages described as safe by United States Cybersecurity and Infrastructure Security Agency, et. al. \cite{cybersecurity2023case}, the unsafe code can be written in the language itself and hidden. For other languages declared to be safe, the unsafe code can be written in other languages and loaded separately. This also hides the unsafe code.

In the past and currently, decisions are and were made to value ease of using unsafe code. These decisions were made to overcome potential or real speed restrictions in the languages and for code reuse reasons. Such decisions tend to encourage the use of unsafe code and encourage programmers and engineers to overlook potential code safety issues found in unsafe code. Heartbleed in OpenSSL \cite{heartbleed2020} is an example of reuse of unsafe code causing vulnerability in large numbers of systems. We question whether these speed and reuse decisions are still relevant. 

In a time of increased and increasing connectivity, increased attacks of various kinds, and increased size of the code bases being created, it is time to value safety over speed and reuse. Unintended consequences of unsafe code are used by intruders to gain unwarranted access to gather data and access or damage computing systems. To reduce these unintended consequences, secure by default and an infinite unsafe impedance should be the goal of every language creator and maintainer. Business practices such as UAPs can not guarantee code safety, only reduce the probability that unsafe code exists in any software product.

We also propose that speed improvements can happen within safe code and in hardware that can mitigate the need for using purposefully written code that may be unsafe. We also propose that using any secure by default language, such as Erlang and Elixir, along with business processes that include an Unsafe Acceptance Process (UAP) aid organizations in producing software that is secure by design \cite{secureByDesignShifting}. These applications, however, are not secure by default since it appears there are no secure by default languages that have infinite unsafe impedance.

Further research can and should be done to expand the assessment of commonly used programming languages with regard to being fully secure by default, i.e. expanding upon the memory CVEs from Table \ref{tab:memory_cves} used in our assessment. Further research should also be done to define a rubric that can be used to rank languages' unsafe impedance.

% The 'abbrvnat' bibliography style is recommended.

\bibliographystyle{plainnat}

\bibliography{unsafe-impedance-final}%remove this for submission

\end{document}